\documentclass{article}
\usepackage{geometry}                
\usepackage{graphicx}
\usepackage{stmaryrd}
\usepackage{amssymb}
\usepackage{amsthm, bm}
\usepackage{bbm}
\usepackage{amsmath, cancel, centernot }
\usepackage[mathscr]{eucal}
\usepackage{mathtools}
\usepackage{epstopdf}
\usepackage{mathrsfs}
\usepackage{newtxmath}

\title{Solving the Mysteries of Quantum Mechanics: Why Nature Abhors a Continuum}
\author{Tim Palmer\\ Department of Physics, University of Oxford, UK\\
tim.palmer@physics.ox.ac.uk}
\date{\today}                                          
\makeatletter
\newcommand\be{\@ifstar{\[}{\begin{equation}}}
\newcommand\ee{\@ifstar{\]}{\end{equation}}}
\newcommand\bp{\begin{pmatrix}}
\newcommand\ep{\end{pmatrix}}

\newtheorem*{theorem*}{Theorem}

\makeatother
\begin{document}
\bibliographystyle{plain}
\maketitle

\abstract{Feynman famously asserted that interference is the only real mystery in quantum mechanics (QM). It is concluded that the reason for this mystery, and thereby the related mysteries of complementarity, non-commutativity of observables, the uncertainty principle and violation of Bell's equality, is that the axioms of QM depend vitally on the continuum nature of Hilbert Space, deemed unphysical. We develop a theory of quantum physics - Rational Quantum Mechanics (RaQM) - in which Hilbert Space is gravitationally discretised. The key to solving the mysteries of QM in RaQM is a number-theoretic property of the cosine function, concealed in QM when angles range over the continuum. This number-theoretic property describes mathematically the utter indivisibility of the quantum world and implies that the laws of physics are profoundly holistic. We contrast holism with nonlocality. In theories which embrace the continuum, the violation of Bell's inequality requires the laws of physics to be either nonlocal or not realistic; both incomprehensible concepts. By contrast, holism, as embodied in Mach's Principle or in the fractal geometry of a chaotic attractor, is neither incomprehensible nor unphysical. As part of this, we solve the deepest mystery of all; why nature makes use of complex numbers.}


\section{Introduction}

In his iconic Lectures on Quantum Mechanics (QM) \cite{Feynman} Feynman describes the mystery of single-particle quantum interference: a phenomenon which `is impossible, \emph{absolutely} impossible, to explain in any classical way, and which has in it the heart of quantum mechanics.'  Feynman believed the phenomenon is so central to QM that `in reality it contains the \emph{only} mystery'. To describe this phenomenon mathematically, QM makes use of the field $\mathbb C$ of complex numbers. Which rather begs a yet-deeper question. Why does QM use an extension $\mathbb C$ of $\mathbb R$ constructed by adjoining the quantity $i$ which, precisely because it is so incomprehensible, can only be defined by axiom. This seems nothing short of miraculous. 

Here we show that the \emph{mystery} of interference (and hence complementarity, non-commutativity of observables, the uncertainty principle and violation of Bell's inequality) is due to QM's dependence on the \emph{continuum} of complex Hilbert Space and hence of $\mathbb C$.  We deem such dependence to be unphysical - a perspective summarised in the title of this paper. Hilbert himself believed that infinity (and hence the infinitesimal) was nowhere to be found in physics \cite{Ellis:2018}, and John Wheeler \cite{Wheeler} (Feynman's PhD supervisor) believed strongly that the continuum of Hilbert Space obscured the information-theoretic nature of the quantum state (thus frustrating his `It from Bit' programme).  Here it is shown that the continuum does more than obscure information, it creates mysteries that can only be solved by developing theories which eschew the continuum. Motivated by these considerations, the author has developed a theory of quantum physics - `Rational Quantum Mechanics' (RaQM) in which the Riemann Sphere, and hence complex Hilbert Space, is discretised. Importantly, as part of this discretisation process, $i$ can be defined constructively as a permutation/negation operator, and not as some incomprehensible object introduced by axiom. 

In RaQM, the discretisation scale is set by gravity. Because gravity is so weak, QM and RaQM are experimentally indistinguishable in many situations. However, an experimental test of RaQM against QM, potentially performable in as few as 5 years, has been proposed using quantum computers  \cite{Palmer:2026b}. QM is the continuum limit of RaQM when $G=0$. This limit is singular \cite{Berry}, implying that the theoretical structure underpinning RaQM is distinct from that of QM. In particular, the theoretical structure of RaQM is underpinned by number theory. By exploiting a number-theoretic property of the cosine function, a property totally concealed in QM where angles range over the continuum, RaQM solves the mystery of interference, and in so doing the related mysteries mentioned above. All this confirms Feynman's assertion that interference is indeed the only real mystery (other claims \cite{Spekkens:2007} notwithstanding -see Section \ref{conclusion}).
 
In solving these mysteries, RaQM finally provides a deep reason why nature makes use of complex numbers. It is to be able to express mathematically (in ways which would be impossible otherwise), the \emph{utter indivisibility} of the quantum world. No additional metaphysical baggage is needed. We already see this with single particle states. Entanglement merely adds to this sense of indivisibility, but not fundamentally so. This points clearly to the  profoundly holistic  - but importantly not nonlocal - nature of the laws of physics. However, there is an important point to be made here. We do not see this holistic structure by invoking the continuum field $\mathbb C$, but instead by invoking a very specific discrete set of complex numbers where, as mentioned, $i$ is defined constructively axiomatically. By contrast, in theories which embrace the continuum, whether $\mathbb C$ or $\mathbb R$, Bell's Theorem can only be interpreted in terms of a violation of locality or realism - notions that manifestly make no physical sense at all. To illustrate two pertinent examples of holism without nonlocality, consider Mach's Principle (that inertia here is due to mass there) and the fractal geometry of a chaotic attractor in state space. 

Section \ref{riemann} discusses the discretisation of the Riemann Sphere, with technical details relegated to the Appendix. Section \ref{raqm} describes the key ideas in RaQM. Section \ref{mysteries} attempts to solve the mysteries of QM in RaQM: starting with single-particle interference and then complementarity, the uncertainty principle, non-commutativity of observables and violation of Bell's inequality. In all cases our understanding of these mysteries boils down to the number-theoretic properties of the cosine function, obscured in QM when angles range over the continuum. In Section \ref{complex} we discuss the mystery of why nature uses complex numbers, and concluding remarks (e.g. implications for synthesising quantum and gravitational physics) are made in the Section \ref{conclusion}. 

\section{Discretised Riemann Sphere and Niven's Theorem}
\label{riemann}

The mathematical basis for RaQM is a discretisation of the Riemann Sphere of extended complex numbers $\mathbb C \cup \{\infty\}$. The discretisation is described in detail in the Appendix. Here we summarise the key properties of the discretisation:
\begin{itemize}
\item $\sqrt{-1}$ is not introduced axiomatically, but is instead defined constructively (and therefore comprehensibly) as a permutation/negation operator (PNO) acting on a 2-element bit string $\{a_1, a_2\}$ where $a_i \in \{1,-1\}$. Specifically $i\{a_1,a_2\}=\{-a_2,a_1\}$, whence $i^2\{a_1,a_2\}=-\{a_1,a_2\}$, where $-\{a_1, a_2\}=\{-a_1, -a_2\}$. At its coarsest discretisation $L=2$, the discretised Riemann sphere comprises four points on a single great circle, representing the four 2-bit strings $\{1,1\}, \{-1,1\}, \{-1,-1\}$ and $\{-1,-1\}$- see Fig \ref{Riemann1} in the Appendix
\item Based on self-similar quaternionic/spinorial concepts (see Figs \ref{Riemann1} and \ref{Riemann2} in the Appendix), the discretised Riemann sphere at some discretisation level $L$ is described in terms of $L$-bit strings at points 
\begin{equation}
\label{rat}
\cos^2 \frac{\theta}{2}= \frac{m}{L} \in \mathbb Q ; \ \ \ \ \frac{\phi}{2\pi}= \frac{n}{L} \in \mathbb Q.
\end{equation}
where $\theta$ denotes co-latitude and $\phi$ denotes longitude on the sphere, $L \in \mathbb N$ defines the degree of granularity of the discretised Riemann Sphere (and hence Hilbert Space) and $0 \le m, n \le L$ are whole numbers. For qubits inside quantum computers, it has been estimated that $L \approx 10^{100}$ if discretisation is due to gravitational effects \cite{Palmer:2026b}. 
\item Like $i$ with $L=2$, the `complex numbers' $\boldsymbol{z}$ on the discretised Riemann Sphere are PNOs acting on the $L$-bit string $\{1,1,\ldots, 1\}$ and defined at points satisfying (\ref{rat}). The fraction of $1$s in an $L$-bit string at the rational point $(\theta, \phi)$ is equal to $\cos^2 (\theta/2)$. $L$-bit strings at rational points $(\theta, \phi_1)$, $(\theta, \phi_2)$ are related by cyclic permutations. 
\end{itemize}

The mathematical key which will allow us to unlock the mysteries of QM is a number theoretic property the cosine function which refer to as Niven's Theorem \cite{Niven} \cite{Jahnel:2005}:
the only values $0 \le \phi < 2\pi$ where $\cos^2(\phi/2)$ - and hence $\cos \phi$ - and $\phi/2\pi$ are simultaneously rational are:
\be
 \phi =0, \frac{\pi}{3},  \frac{\pi}{2},  \frac{2\pi}{3},  \pi, \frac{4\pi}{3},  \frac{3\pi}{2},  \frac{5\pi}{3}.
 \ee
Hence, unless $\phi$ is a multiple of $60^\circ$ or $90^\circ$ \emph{exactly},  $\cos \phi$ and $\phi/2\pi$ cannot be simultaneously rational. In particular, the cosines of angles like $0.0000001^\circ$ and $60.0000001^\circ$ are irrational. In a small neighbourhood of $0^\circ$ comprising $10^{100}$ rational angles (what below we refer to as a nominal direction of $0^\circ$), the likelihood of choosing an angle whose cosine is rational is equal to one part in $10^{100}$, i.e., utterly negligible. 

\begin{figure}
\centering
\includegraphics[scale=0.4]{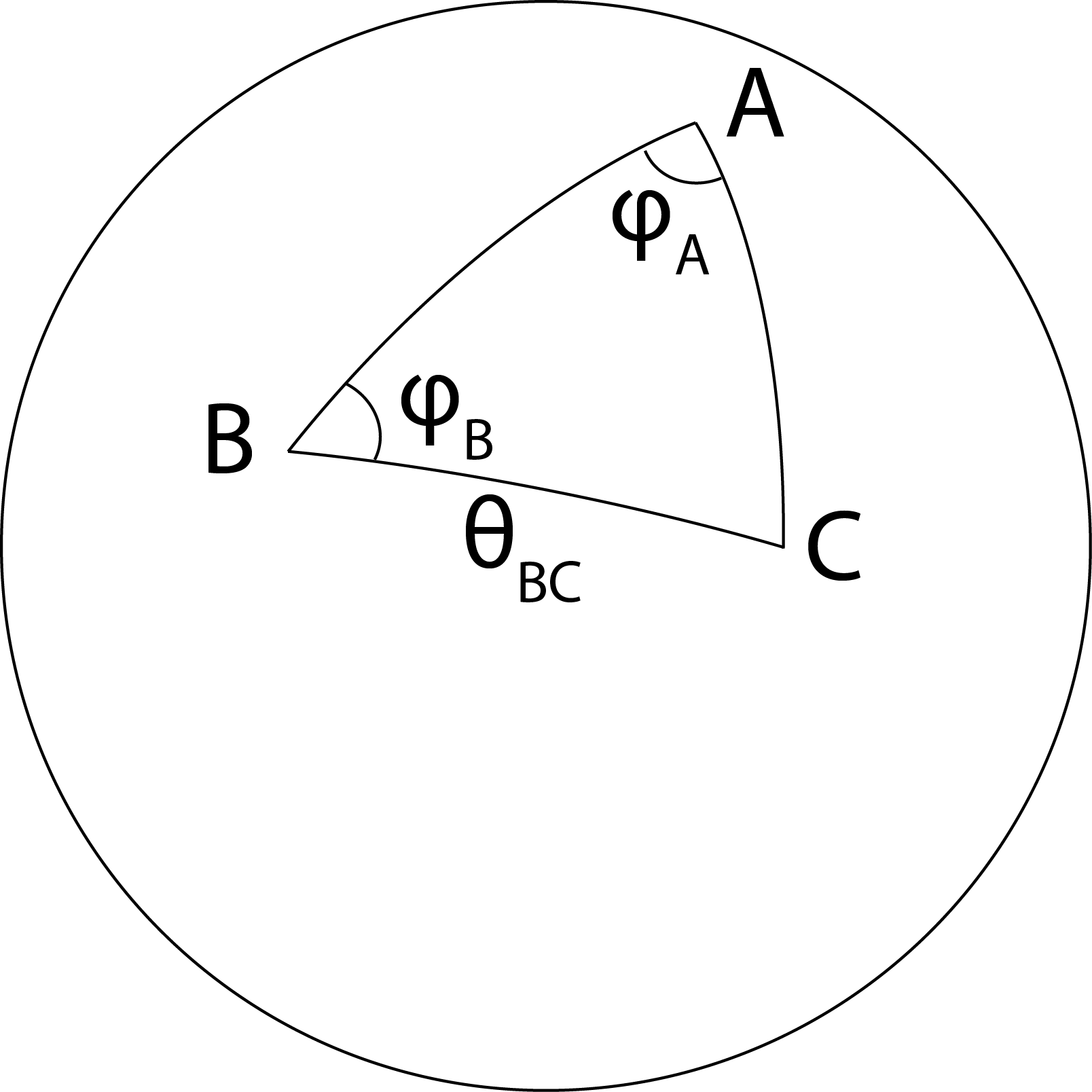}
\caption{\emph{An impossible spherical triangle, where the cosine of the angular length of each side of the triangle on the unit sphere is rational, and the internal angles are rational in degrees. The Impossible Triangle Corollary (to Niven's Theorem) provides the basis for understanding the mysteries of QM.}}
\label{triangle}
\end{figure}

In practice, the examples discussed in Section \ref{mysteries} make use of a simple corollary to Niven's Theorem referred to as the Impossible Triangle Corollary (ITC). Consider a spherical triangle $\triangle ABC$ on the unit sphere with vertices $A$, $B$ and $C$ (see Fig \ref{triangle}). If the angular lengths of the three sides are written $\theta_{AB}$, $\theta_{BC}$ and $\theta_{AC}$, then the cosine rule applied to $\triangle ABC$ can be written
\be
\label{sp}
\cos \theta_{AC}= \cos \theta_{AB} \cos \theta_{BC} + \sin \theta_{AB} \sin \theta_{BC} \cos \phi_C
\ee
Can all three of these angular lengths can have rational cosines, if the interior angles are themselves rational (in degrees)? Let us assume that they can. That is to say, assume that both the LHS and the first term on the RHS of (\ref{sp}) is rational. Then the second term on the RHS of (\ref{sp}) must also be rational. Squaring, this implies that 
\be
(1-\cos^2 \theta_{AB})(1- \cos^2 \theta_{BC}) \cos^2 \phi_C \in \mathbb Q
\ee
and hence $\cos^2 \phi_C \in \mathbb Q$ and therefore $\cos 2\phi_C \in \mathbb Q$. However, apart from the very sparse exceptions to Niven`s Theorem, this cannot be so if $\phi$ itself is a rational multiple of $2\pi$. As discussed below, this number-theoretic result will be used to show that the mystery of entanglement as per Bell's Theorem is really no different to the mystery of single-particle interference. 

\section{RaQM}
\label{raqm}

\subsection{Single-Qubit Systems}

In RaQM, the Schr\"{o}dinger evolution is not modified. Rather, we restrict the Dirac - von Neumann axiom so that, instead of a unitarily evolved qubit state 
\begin{equation}
\label{qubit3}
|\psi(\theta,\phi) \rangle= \cos \frac{\theta}{2} |1\rangle + e^{i \phi} \sin\frac{\theta}{2} |-1\rangle
\end{equation}
being defined in any basis of eigenvectors of some Hermitian observable, it is only defined in bases $\{|1\rangle, |-1\rangle\}$ where (\ref{rat}) holds. 

Relative to a basis where (\ref{rat}) does not hold (e.g. if squared amplitudes or phases in degrees are irrational reals), the Hilbert state (\ref{qubit3}) is undefined. As discussed below, this restriction of the Dirac-von Neumann axiom makes mathematical sense using the $p$-adic metric to describe distances between rationals. With respect to this metric, the irrational reals are indefinitely far away, and hence undefined. 

The key to RaQM's representation of $|\psi\rangle$ lies in the representation of points on the discretised Riemann Sphere in terms of length-$L$ bit string. Specifically, if $\theta$, $\phi$ satisfy  (\ref{rat}) then
\begin{align}
\label{bits}
|\psi(\theta, \phi) \equiv \{\underbrace{\ 1,\; \ \; 1,\ \; 1,\ \ \dots \ \ 1,\ \; 1,\ \; 1,\ \ -1,-1,-1, \ \ldots \ ,-1,-1,-1}_{\theta, \phi}\}\ \ \ \bmod \xi 
\end{align}
comprising the bits `1' and '-1'. As before, $\cos^2 \theta/2 = m/L$ equals the fraction of 1s in the bit string and $\phi/2\pi=n/L$ encodes $n$ cyclic permutations of the bit string, corresponding to a complex phase rotation in QM. 

The Hilbert state in QM is invariant under a global phase. Corresponding to this, the state defined in (\ref{bits}) is an equivalence class of bit strings under general permutations $\xi$. In this sense, (\ref{bits}) describes an unordered ensemble of $L$ possible binary measurement outcomes (described by symbolic labels \cite{Schwinger}).with frequencies consistent with Born's rule.

However, unlike in QM, global phase plays a key role in RaQM (transforming it from a probabilistic to a deterministic theory). In particular, for some specific $\xi$ we write
\begin{align}
\label{bits2}
|\psi(\theta, \phi) \rangle_\xi = \xi  \{\underbrace{\ 1,\; \ \; 1,\ \; 1,\ \ \dots \ \ 1,\ \; 1,\ \; 1,\ \ -1,-1,-1, \ \ldots \ ,-1,-1,-1}_{\theta, \phi}\}
\end{align}
describing a specific ordered ensemble of bits. Indeed, (\ref{bits2}) describes the difference between two bitwise complementary integers in base 2. For example, with $L=4$ we write 
\be
\label{bits3}
|\psi(\frac{\pi}{2}, 0) \rangle_\xi = \{1,-1, -1, 1\} = 1001.-0110.
\ee
This integer representation describes the state of \emph{a specific} quantum system e.g. a specific photon, in a rational basis. As such, $\xi$ can be thought of as a hidden variable (hidden to us), describing the relationship between the quantum system and the rest of the universe, and fixed the moment the quantum system is created. 

$\xi$ is key to describing the outcome $\pm 1$ of a measurement on a specific individual quantum system. In particular, in \cite{Palmer:2026b}, measurement is described in terms of a chaotic bit shift, where the integer representation is divided by 2 every Planck time-step. In the $L=4$ example (\ref{bits3}) above,  
\be
1001.-0110. \mapsto 100.-011. \mapsto 10. -01. \mapsto 1.
\ee
In this representation, the integers $1001.$ and $0110.$ are, as discussed above, considered relative to the 2-adic metric. As such, the bits describe pieces of a Cantor Set $\mathscr C$, where the further to the left a bit is in the integer, the smaller the piece of $\mathscr C$ it labels (in contradistinction to the more familiar Euclidean metric where the further to the left a bit is positioned in the integer, the larger the interval of the line it represents): see \cite{Palmer:2026b} for details. As such, dividing by 2 describes a map from one iteration of $\mathscr C$ to one larger. In the example above, state reduction occurs over $L-1=3$ Planck times. When $L=10^{100}$ say, state reduction would take longer than the age of the universe. Estimates for $L$ associated with qubits in a quantum computer are discussed in \cite{Palmer:2026b}. As such RaQM should be testable against QM in a few years, if quantum tech roadmaps are to be believed. 

The use of the 2-adic metric here implies that bases where (\ref{rat}) is violated because the squared amplitudes and/or phases are irrational real numbers, are mathematically undefined, being infinitely far from the bases where (\ref{rat}) is satisfied. By contrast relative to a Euclidean metric, the irrational bases would be close to the rational bases and RaQM would be considered a very fine-tuned theory. The use of the 2-adic metric here makes clear why we cannot think of RaQM as a classical theory. 

Since $\xi$ is hidden, we can for all practical purposes think of the measurement outcome as randomly chosen (c.f. the Copenhagen Interpretation of QM). However, it is important to note that the notion of randomness here is not fundamental, any more than rolling a dice is fundamentally random. Effectively, by not `modding over' global phase, $\xi$ allows a deterministic description of single-particle states, not possible in QM. 

\subsection{Multipartite Systems}

In RaQM, a quantum system comprising $N$ entangled qubits is represented as $N$ correlated length-$L$ bit strings, where the same permutation $\xi$ applies to each string (consistent with $\xi$ as a global phase in QM). As discussed in \cite{Palmer:2026b}, providing $L$ is large enough, there are as many of degrees of freedom in these $N$ bit strings, as in an $N$ qubit system in QM ($2^{N+1}-2$). In this paper, we will only be concerned with $N=1$ or $N=2$. A general, explicitly normalised, $N=2$-qubit state in QM can be written
\begin{equation}
\label{threeraqm}
\begin{split}
\begin{array}{cc}
|\psi_{AB}\rangle=\underbrace{\cos \frac{\theta_1}{2}|1_A\rangle}_A\  
\times\  (\underbrace{\cos\frac{\theta_2}{2}|1_B\rangle+e^{i \phi_2} \sin\frac{\theta_2}{2}|-1_B\rangle}_B)+ \\
\ \ \ \ \ \ \ \ \ \ \  \ \  \ \ \  \ \ \  \ \  +\underbrace{e^{i \phi_2}\sin \frac{\theta_1}{2}|-1_A\rangle}_A
\times (\underbrace{\cos\frac{\theta_3}{2}|1_B\rangle+e^{i \phi_3} \sin\frac{\theta_3}{2}|-1_B\rangle}_B) 
\end{array}
\end{split}
\end{equation}
If the individual amplitudes and phases in (\ref{threeraqm}) satisfy (\ref{rat}), then in RaQM $|\psi_{AB}\rangle$ is the equivalence class of 2 length-$L$ bit strings
\be
\label{twobitstrings}
|\psi_{AB}\rangle \equiv \left\{
\begin{array} {c}
\{\underbrace{\ 1,\; \ \; 1,\ \; 1,\ \ \dots \ \ 1,\ \; 1,\ \; 1,\ \ -1,-1,-1, \ \ldots \ ,-1,-1,-1}_{\theta_1, \phi_1}\}\ \ \ \bmod \xi  \\
\{\underbrace{1,1,\; \ldots , 1, \ -1, -1,\ldots,-1}_{\theta_2, \phi_2} \ \underbrace{\ 1,1, \ldots ,\ \ 1,\ -1,-1 \ldots \ ,-1}_{\theta_3, \phi_3}\}\ \ \ \bmod \xi 
\end{array}
\right.
\ee
with 6 degrees of freedom, as in QM. The top string represents the A qubit, and the bottom representing the B qubit, where $\theta_i$ and $\phi_i$ satisfy the rationality conditions (\ref{rat}). Hence, for example, $\cos^2 \theta_2/2$ denotes the fraction of $1$ bits in the B bit string which correspond to 1 bits in the A bit string, and $\phi_2$ represents a cyclic permutation of specific bits in the B bit string, as shown. Again, $\xi$ corresponds to a permutation common to both bit strings, corresponding to some global phase. As such, in RaQM, the quantum state for a specific pair of qubits can be written  
\be
\label{twobitsxi}
|\psi_{AB}\rangle_\xi=\left\{
\begin{array} {c}
\xi \{\underbrace{\ 1,\; \ \; 1,\ \; 1,\ \ \dots \ \ 1,\ \; 1,\ \; 1,\ \ -1,-1,-1, \ \ldots \ ,-1,-1,-1}_{\theta_1, \phi_1}\}  \\
\xi \{\underbrace{1,1,\; \ldots , 1, \ -1, -1,\ldots,-1}_{\theta_2, \phi_2} \ \underbrace{\ 1,1, \ldots ,\ \ 1,\ -1,-1 \ldots \ ,-1}_{\theta_3, \phi_3}\}
\end{array}
\right.
\ee

For Bell's Theorem below, we focus on the QM singlet state
\be
\label{singlet}
|\psi^{\mathrm{singlet}}_{AB}\rangle=\frac{1}{\sqrt 2}\left (|\ 1_A\rangle |-1_B\rangle-|-1_A\rangle|\ 1_B\rangle \right)
 \ee
where A and B denote Alice and Bob's qubit respectively. With respect to a basis where Alice's measuring apparatus is oriented an an angle $\theta_{AB}$ to Bob's, (\ref{singlet}) can be written
\be
\label{singlet2}
|\psi^{\mathrm{singlet}}_{AB}\rangle=\cos\frac{\theta_{AB}}{2} \left ( \frac{|\ 1_A\rangle |-1_B\rangle- |-1_A\rangle |\ 1_B\rangle}{\sqrt 2} \right )+ \sin\frac{\theta_{AB}}{2} \left ( \frac{|\ 1_A\rangle |\ 1_B\rangle - |-1_A\rangle |-1_B\rangle}{\sqrt 2} \right ) 
\ee
Providing $\theta_{AB}$ satisfies (\ref{rat}), for a specific run of a Bell experiment, the singlet state (\ref{singlet2}) in RaQM is represented by the 2 length-$L$ bit strings
\be
\label{bitstrings}
|\psi^{\mathrm{singlet}}_{AB}\rangle_\xi=\left\{
\begin{array} {c}
\xi \{\underbrace{\ 1,\; \ \; 1,\ \; 1,\ \ \dots \ \ 1,\ \; 1,\ \; 1,\ \ -1,-1,-1, \ \ldots \ ,-1,-1,-1}_{\frac{\pi}{2}, 0}\} \\
\xi \{\underbrace{1,1,\; \ldots , 1, \ -1, -1,\ldots,-1,}_{\pi-\theta_{AB}, 0} \ \underbrace{\ 1,1, \ldots ,\ \ 1,\ -1,-1 \ldots \ ,-1}_{\theta_{AB}, 0}\} 
\end{array}
\right.
\ee
As before, the measurement outcomes is determined by $\xi$, fixed at the time the entangled particles leave their source region, and hence is `known' to both qubits. For us, for whom $\xi$ is hidden, the measurement outcomes are, for all practical purposes, determined by a random choice of some $M$th pair of bits, one from the top bit string and one from the bottom. In particular, when $\theta_{AB}=0$, then if Alice measures $+1$, Bob will certainly measure $-1$ and \emph{vice versa}. 

Following EPR \cite{EPR} and Bell \cite{Bell:1964}, we will say that a theory is EPR/Bell nonlocal if, for a particular run of a Bell experiment, Alice's measurement outcome can depend on Bob's measurement setting, and \emph{vice versa}. It is easy to see that RaQM is not EPR/Bell nonlocal. Suppose, for a particular entangled particle pair, Alice and Bob choose nominal  measurement orientations $\boldsymbol A_{\mathrm{nom}}$ and Bob $\boldsymbol B_{\mathrm{nom}}$, where the exact settings $\boldsymbol A$ and $\boldsymbol B$ are consistent with these nominal settings and $\boldsymbol A \cdot \boldsymbol B$ satisfies (\ref{rat}). If Bob is a free agent, then he could have chosen any other nominal direction $\boldsymbol B'_{\mathrm{nom}}$ providing there exist an exact setting $\boldsymbol B'$ where $\boldsymbol A \cdot \boldsymbol B'$ also satisfies (\ref{rat}). From (\ref{bitstrings}), if Bob counterfactually changes his measurement orientation from $\boldsymbol B_{\mathrm{nom}}$ to $\boldsymbol B'_{\mathrm{nom}}$, Alice's bit string does not change. In particular, Alice's $M(\xi)$th bit, corresponding her measurement outcome, does not change. 

We can similarly argue that if Alice changes her mind, it won't affect Bob's measurement outcome. This is because, given $\xi$, there exists a corresponding $\xi'$ such that 
\be
\label{bitstrings}
|\psi^{\mathrm{singlet}}_{AB}\rangle_{\xi}=\left\{
\begin{array} {c}
\xi' \{\underbrace{1,1,\; \ldots , 1, \ -1, -1,\ldots,-1,}_{\pi-\theta_{AB}, 0} \ \underbrace{\ 1,1, \ldots ,\ \ 1,\ -1,-1 \ldots \ ,-1}_{\theta_{AB}, 0}\} \\
\xi' \{\underbrace{\ 1,\; \ \; 1,\ \; 1,\ \ \dots \ \ 1,\ \; 1,\ \; 1,\ \ -1,-1,-1, \ \ldots \ ,-1,-1,-1}_{\frac{\pi}{2}, 0}\}
\end{array}
\right.
\ee

It is important to note that the rationality restrictions on $\boldsymbol A \cdot \boldsymbol B$ and $\boldsymbol A \cdot \boldsymbol B'$ do not impose any constraint on the ability of experimenters to choose their measurement settings freely. When Alice chooses some nominal direction $\boldsymbol A_{\mathrm{nom}}$, and Bob some nominal direction $\boldsymbol B_{\mathrm{nom}}$, their choice defines, not a pair of single points on the celestial sphere, but a pair of small neighbourhoods of points. Within these neighbourhoods there will certainly will exist exact measurement orientations $\boldsymbol A$ and $\boldsymbol B$ such that $\boldsymbol A \cdot \boldsymbol B \in \mathbb Q$. If Bob counterfactually changes his mind, $\boldsymbol B'_{\mathrm{nom}}$ again denotes a small neighbourhood of points on the celestial sphere, and again there will exist exact pairs of points, one from the neighbourhood defined by $\boldsymbol A_{\mathrm{nom}}$ and one from the neighbourhood defined by $\boldsymbol B'_{\mathrm{nom}}$, which satisfy the rationality condition $\boldsymbol A \cdot \boldsymbol B' \in \mathbb Q$. 

The key question to be asked is whether RaQM satisfies Bell's inequality. In Section \ref{bell} we show that the rationality constraints on the exact measurement settings prevent Bell's inequality from being derived in the usual way (locality and realism notwithstanding). The argument hinges around an extension of the discussion above, where we are required to consider not one but two simultaneous counterfactuals. In such a situation, the rationality conditions have some bite, as shown by ITC. 

\section{ Solving the Mysteries of QM}
\label{mysteries}

\subsection{Quantum Interference}
\label{interference}

The essence of interference in QM is described by the Mach-Zehnder interferometer. A single photon in state 
$|-1\rangle =\big(\begin{smallmatrix}
0 \\
1
\end{smallmatrix} \big )$
passes through a beamsplitter. A phase shifter is positioned in the lower arm of the interferometer, resulting in the photon being in the superposed state
\be
\label{during}
\begin{pmatrix}
1&0\\
0&e^{i \phi}
\end{pmatrix}
\frac{1}{\sqrt 2}
\begin{pmatrix}
1&1\\
1&-1
\end{pmatrix}
\begin{pmatrix}
0\\
1
\end{pmatrix}
=
\frac{1}{\sqrt 2}
\begin{pmatrix}
1\\
e^{i \phi}
\end{pmatrix}
=\frac{1}{\sqrt 2}( |1\rangle + e^{i\phi} |-1\rangle
\ee
After passing through a second beamsplitter the state becomes
\be
\label{after}
\frac{1}{\sqrt 2}
\begin{pmatrix}
1&1\\
1&-1
\end{pmatrix}
\begin{pmatrix}
1\\
e^{i \phi}
\end{pmatrix}=
\frac{1}{2}
\begin{pmatrix}
1+e^{i \phi}\\
1-e^{i \phi}
\end{pmatrix}
= \sin \frac{\phi}{2} |1\rangle -i \cos \frac{\phi}{2} |-1\rangle
\ee
where we use the complex-number identities,
\begin{align}
\label{identity}
2 \cos \frac{\phi_A-\phi_B}{2}e^{i\frac{\phi_A+\phi_B}{2}}&=e^{i \phi_A}+e^{i \phi_B} \nonumber \\
2i \sin \frac{\phi_A-\phi_B}{2}e^{i\frac{\phi_A+\phi_B}{2}}&=e^{i \phi_A}-e^{i \phi_B},
\end{align}
 a consequence of Euler's formula. The evolved state $\sin \frac{\phi}{2} |1\rangle -i \cos \frac{\phi}{2} |-1\rangle$ describes constructive/destructive wave interference as $\phi$ varies from $0^\circ$ to $180^\circ$.  
 
We now see how RaQM can help solve an otherwise incomprehensible mystery. For the state in (\ref{after}) to satisfy (\ref{rat}) it is necessary that $\cos^2 (\phi/2) \in \mathbb Q$, i.e. $\cos \phi \in \mathbb Q$. However, for the state in (\ref{during}) to satisfy (\ref{rat}), it is necessary that $\phi/2\pi \in \mathbb Q$. By Niven's Theorem, these conditions virtually always incompatible. 

Of course, this doesn't mean that the Hilbert State is undefined when the quantum system is passing through the interferometer. Rather, it means that, if $\cos \phi \in \mathbb Q$, then $|\psi\rangle$ is undefined in a basis where the upper and lower arms of the interferometer correspond to the independent orthogonal basis vectors.  Clearly it will be possible to find rotations $\{|1'\rangle, |-1'\rangle\}$ of $\{|1\rangle, |-1\rangle\}$ where the photon state does satisfy (\ref{rat}). But now the independent orthogonal basis elements necessarily correspond to worlds which explicitly recognise the combined existence of the upper and lower arms. That is to say, the state of the photon (inside the interferometer) can only be defined in bases which recognise the holistic nature of the interferometer. 

As such, asking questions of RaQM which do not recognise this holistic nature (e.g., `did the particle go along the upper arm or the lower arm?') do not require convoluted metaphysical discussion. Rather, they simply and straightforwardly have no mathematical answer in RaQM: to answer the question requires projecting the quantum state into a basis where the rationality conditions cannot be satisfied (no matter how fine is the discretisation of Hilbert Space). Only in the unphysical continuum limit (i.e. in QM) is the mathematical answer to such a question not ill posed. But now we do require convoluted metaphysical gymnastics to answer the question, because any comprehensible answer is impossible. Or we simply evade answering the question (`shut up and calculate'). For example, in discussing single electrons in the famous two-slit experiment, Feynman asks `Is it true, or is it \emph{not} true, that the electron either goes through hole 1 or it goes through hole 2?'  He concludes: `The only answer that can be given is that we have found from experiment there is a certain way we have to think in order we do not get into inconsistencies. What we must say $\ldots$ is the following. If one looks at the holes $\ldots$ then one \emph{can} say that [the electron] goes either through hole 1 or hole 2. \emph{But}, when one does \emph{not} try to tell which way the electron goes, when there is nothing in the experiment to disturb the electrons, then one may \emph{not} say that an electron goes through either hole 1 or hole 2.'  Of course, this is exactly right. However, finally, RaQM can explain \emph{why} it is right. RaQM solves the deep mystery of quantum interference without resort to incomprehensible metaphysical baggage. 

The notion that the laws of physics are holistic (but, importantly, not nonlocal) also emerges from our analysis of Bell's inequality in Section \ref{bell}. But no new elements are required to analyse Bell's inequality, our conclusion is also a simple consequence of Niven's Theorem. And, unlike nonlocality, there is nothing incomprehensible about holistic laws of physics (see Section \ref{conclusion}).   

When viewed in this way, phenomena like the Aharanov-Bohm effect become straightforward to understand. It is not that the photon is locally affected by the vector potential of the magnetic field, nor that the magnetic field has some nonlocal effect on the photon, but rather than the photon state is only defined in bases whose individual orthogonal basis vectors describe worlds which encompass the two arms of the interferometer and thereby the magnetic solenoid enclosed within these arms. 

Another mystery that is straightforwardly solved by RaQM is Wheeler's delayed-choice experiment - where the experimenter only decides to leave the second half-silvered mirror in place, or remove it, after the photon has entered the interferometer. The mystery concerns how the photon knows whether to behave as a wave or a particle inside the interferometer if the experimenter hasn't yet decided what type of measurement to make.  The answer from RaQM comes in two parts. The first is that the state of the photon in RaQM is not `a wave' or a `a particle', but is an $L$-bit string which shows either wave-like or particle-like characteristics according to whether $\cos \phi$ is rational, or $\phi$ is rational, respectively. The second is that $\cos \phi$ will be rational if the second half-silvered mirror is in place, and $\phi$ is rational if the second half-silvered mirror is removed. This is entirely consistent with the holistic definition of the photon state in RaQM. Hence if the second half silvered mirror is initially in place, the $L$-bit state of the photon will be consistent with rational $\cos \phi$.  If the experimenter removes the second half-silvered mirror, the $L$-bit state of the photon will transform to one consistent with $\phi$ being rational. There is no `spooky action at a distance' here because, as discussed, the quantum state is not localisable to one arm of the interferometer or the other. It is only localisable to within the interferometer. 
 
The identities (\ref{identity}) contain the heart of the matter under discussion. If the rationality conditions are met for the LHS, they cannot be met for the individual phases on the RHS. Indeed, the essence of the argument is contained in the ITC. Consider two complex numbers $z_1$ and $z_2$, and their sum and difference $z_1 \pm z_2$ on the Argand plane with origin $O$. We can map the triangle $\triangle Oz_1z_2$ onto the Riemann Sphere. ITC implies that if the angular distances of the corresponding mapped sides $Oz_1$ and $Oz_2$ satisfy (\ref{rat}), and the internal angles are rational, then the angular distances of $O (z_1 \pm z_2)$ do not. Hence, as with all other QM mysteries below, the solution lies in the role of ITC in RaQM.  

\subsection{Complementarity}
\label{complementarity}

The discussion above on interference illustrates wave-particle complementarity. In RaQM, complementarity is merely the expression that in practice (i.e. with overwhelming probability) if $\cos \phi$ is rational (wave-like $L$-bit string), then $\phi$ cannot be rational (particle-like $L$-bit string) and \emph{vice versa}. 

Another example of complementarity in RaQM comes from the second equation in (\ref{identity}). If we write $\phi_1= k(x + \Delta x)$, $\phi_2 = k(x -\Delta x)$, then the second of  (\ref{identity}) becomes the finite-difference equation
\be
\label{pm}
i \frac{\Delta e^{ikx}}{\Delta x} = 2i e^{i kx} \frac{\sin kx}{\Delta x}
\ee
If we assume $kx \in \mathbb Q$ and $\Delta x \in \mathbb Q$, so that the individual terms comprising the LHS of  (\ref{pm}) satisfy the rationality conditions, then the RHS does not, because, with these conditions, $\sin^2 kx \notin \mathbb Q$. Conversely, if the term on the RHS satisfies the rationality conditions, then the individual terms that comprise the finite difference on the LHS do not. 

That is to say, position and momentum are not simultaneously definable. Such analysis starts to probe at what RaQM says about the discretisation of space and time itself. We leave this for another paper. 

\subsection{Uncertainty}
\label{uncertainty}

\begin{figure}
\centering
\includegraphics[scale=0.3]{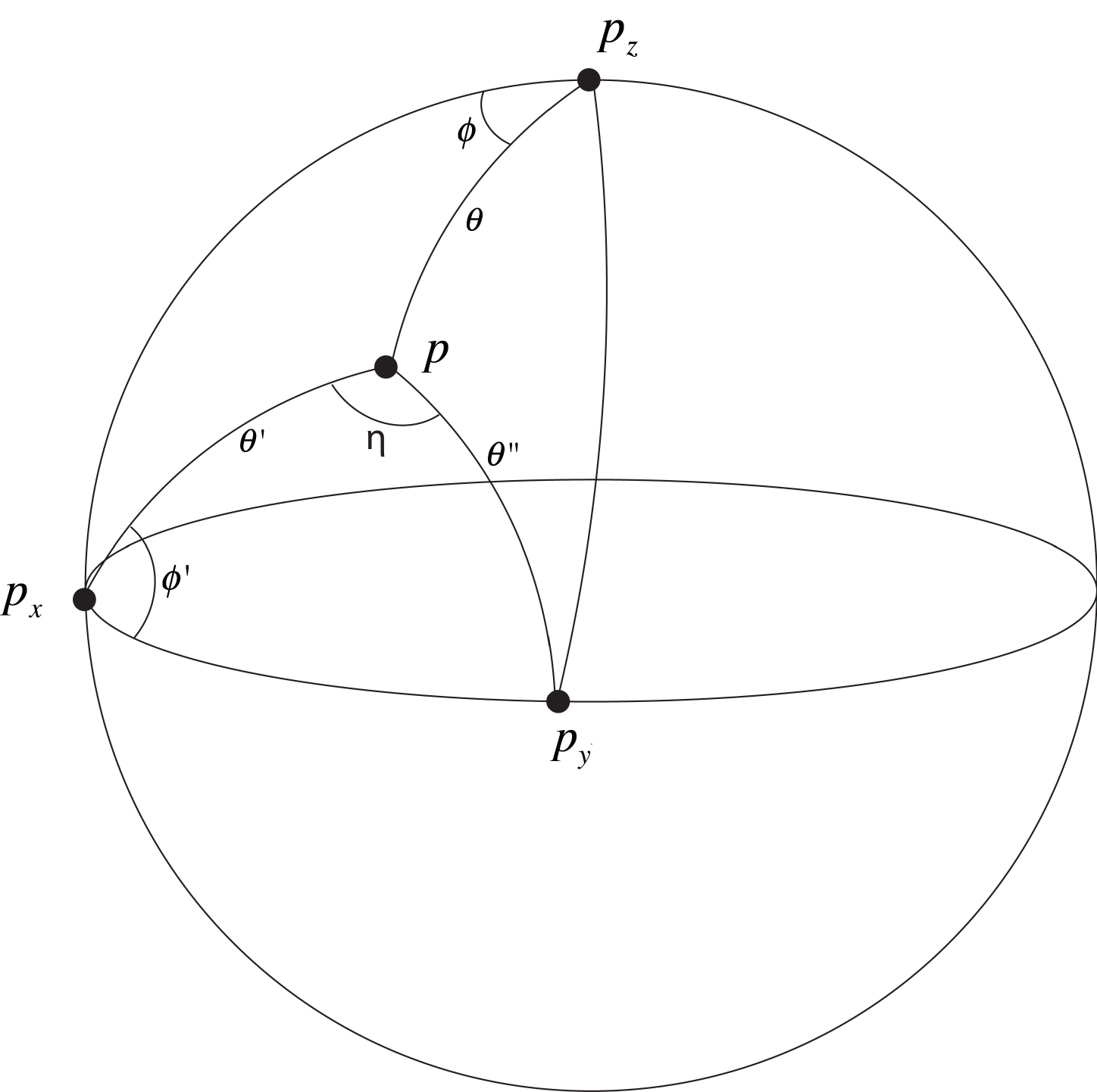}
\caption{\emph{In RaQM, the Uncertainty Principle $\Delta S_x \Delta S_y \ge \frac{\hbar}{2} \overline S_z$  for a single qubit arises from the trigonometry of spherical triangles and the correspondence between points on the discretised Riemann sphere with coordinates satisfying the rationality conditions and corresponding bit strings.}}
\label{uncertainty}
\end{figure}
In RaQM, the Uncertainty Principle for a single qubit arises from a trigonometric inequality on the sphere, together with Niven's Theorem. We then discuss the Uncertainty Principle for multiple qubits. 

Consider a point $p$ on the unit sphere (Fig \ref{uncertainty}) whose colatitude with respect to the three orthogonal poles $p_x$, $p_y$ and $p_z$ is $\theta$, $\theta'$ and $\theta''$ respectively. The internal angles $\phi'$ and $\eta$ are shown on the figure. By the sine rule for spherical triangle $\bigtriangleup pp_xp_y$
\begin{equation}
\frac{\sin \theta''}{\sin \phi'} = \frac{\sin \pi/2}{\sin \eta}=\frac{1}{\sin \eta} 
\end{equation}
Hence
\begin{equation}
\label{unc1}
|\sin \theta''| \ge |\sin \phi'|
\end{equation}
By the cosine rule for spherical triangle $\bigtriangleup pp_xp_z$, 
\begin{equation}
\label{unc2}
\cos \theta = \sin \theta' \sin \phi'
\end{equation}
From (\ref{unc1}) 
\begin{equation}
|\sin \theta'| |\sin \theta''| \ge |\sin \theta'| |\sin \phi'| 
\end{equation}
and using (\ref{unc2})
\begin{equation}
\label{unc3}
|\sin \theta'| |\sin \theta''| \ge |\cos \theta| 
\end{equation}
From RaQM, a bit string at colatitude $\theta$ has a mean value $\mu_{\theta}=\cos\theta$, and standard deviation $\sigma_\theta = \sin \theta$. With this in mind, consider three discretised Bloch spheres, with the north poles oriented at $p_x$, $p_y$ and $p_z$ respectively. Then from (\ref{unc3}),
\begin{equation}
\label{unc4}
\sigma_{\theta'}\; \sigma_{\theta''} \ge |\mu_{\theta}| 
\end{equation}
If instead of $\pm 1$, the bit strings have dimensional values $\pm \hbar/2$ in order that they correspond to physical spin, then (\ref{unc4}) becomes the familiar uncertainty principle for spin qubits
\begin{equation}
\label{unc5}
\Delta S_x\; \Delta S_y \ge \frac{\hbar}{2}\; \overline S_z
\end{equation}

The rationality constraints play a key role here. If, for example, $\cos \theta \in \mathbb Q$ and $\phi/2\pi \in \mathbb Q$, by Niven's Theorem applied to $\bigtriangleup pp_xp_z$, $\cos \theta'$ cannot be rational. Hence it is impossible to know simultaneously, the spin values of a particle with respect to the two directions $p_x$ and $p_z$ (similarly for any two other pairs of directions). 

Let us now derive the Uncertainty Principle for conjugate observables defined on multiple qubits (like position and momentum). First we square (\ref{unc4}) and average over $M$ points $p$ on the discretised Bloch Sphere. Then, from (\ref{unc3}), 
\be
\overline{\sigma^2_{\theta'}\; \sigma^2_{\theta''}} \ge \overline{|\mu_{\theta}|^2}
\ee
Now 
\begin{align}
\overline{\sigma^2_{\theta'}} \ \overline{\sigma^2_{\theta''}} =&\frac{1}{M} (\sigma^2_{\theta'_1}+\sigma^2_{\theta'_2} + \ldots + \sigma^2_{\theta'_M}) 
 \times \frac{1}{M} (\sigma^2_{\theta''_1}+\sigma^2_{\theta''_2} + \ldots + \sigma^2_{\theta''_M}) \nonumber \\
&> \frac{1}{M} ((\sigma^2_{\theta'_1}\sigma^2_{\theta''_1}+\sigma^2_{\theta'_2}\sigma^2_{\theta''_2} + \ldots + \sigma^2_{\theta'_M}\sigma^2_{\theta''_M}) 
= \overline{\sigma^2_{\theta'} \sigma^2_{\theta''}}
\end{align}
Hence, if the $M$ points are uniformly distributed with respect to $\cos \theta$ so that $\overline{|\cos \theta|}=1/2$, then
\be
\sqrt{\overline{\sigma^2_{\theta'}}} \ \sqrt{\overline{\sigma^2_{\theta''}}} \ge \frac{1}{2}
\ee
Multiplying by $\hbar$ gives the standard position/momentum form
\be
\Delta x \Delta p \ge \frac{1}{2} \hbar
\ee
of the Uncertainty Principle. 

\subsection{Non-Commutativity}
\label{noncommutativity}

\begin{figure}
\centering
\includegraphics[scale=0.7]{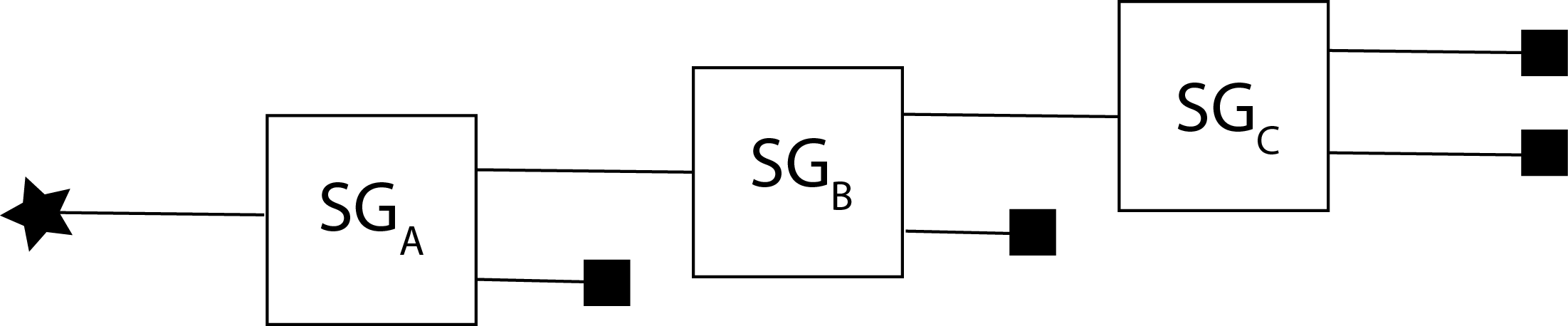}
\caption{\emph{A spin-$1/2$ quantum system is prepared `up' by Stern-Gerlach device $SG_A$. The spin-up output of $SG_A$ is fed into device $SG_B$ and the spin-up output of $SG_B$ is fed into $SG_C$. The experimenter is free to choose the nominal orientations of these apparatuses as they like. However, according to the rationality constraints (\ref{ratSG2}), Niven's Theorem prevents the simultaneous counterfactual world where $SG_B$ and $SG_C$ are swapped from having a well-defined measurement outcome.}}
\label{SG}
\end{figure}

By passing a spin-1/2 particle through a Stern-Gerlach ($SG_A$) device oriented along $\boldsymbol A$, a particle is prepared with spin up relative to some direction $\boldsymbol A$,  (see Fig \ref{SG}). The particle is then passed through a second Stern-Gerlach device ($SG_B$) along direction $\boldsymbol B$. The spin-up output of $SG_B$ is then passed to a third Stern-Gerlach device ($SG_C$) oriented along direction $\boldsymbol C$. Again, $E$ is free willed and orientates the SG devices (to some nominal accuracy) as they like. Let $A$, $B$ and $C$ denote vertices of a spherical triangle $\triangle ABC$ on the unit (celestial) sphere, corresponding to the tips of the three unit vectors $\boldsymbol A$, $\boldsymbol B$ and $\boldsymbol C$ in physical space. See Fig \ref{triangle}. 

We will assume that a given quantum system whose spin is prepared `up' relative to $SG_A$, has spin $S(\lambda, \boldsymbol A, \boldsymbol B)$ coming out of $SG_B$ where $S$ is some deterministic function returning either $+1$ or $-1$ and $\lambda$ is a hidden variable (given by the permutation $\xi$ in Section \ref{raqm}). Similarly, the same quantum system prepared `up' relative to $SG_B$, has spin $S(\lambda, \boldsymbol B, \boldsymbol C)$ coming out of $SG_C$. By definitiion, $\boldsymbol A \cdot \boldsymbol B = \cos \theta_{AB}$ where $\theta_{AB}$ is the angular distance between $A$ and $B$ on the unit sphere, etc.

Although $\boldsymbol A$, $\boldsymbol B$ and $\boldsymbol C$ will be approximately coplanar, they will not be exactly coplanar. Hence $\triangle ABC$ is non-degenerate, but where $\phi_B \approx \pi$.  The rationality constraints (\ref{rat}) imply that 
\be
\label{ratSG2}
\cos \theta_{AB} \in \mathbb Q;\ \ 
\cos \theta_{BC} \in \mathbb Q.
\ee
In addition, the angle $\phi_B$ between the great circles AB and BC must satisfy
\be
\label{ratSG3}
\frac{\phi_B}{2\pi} \in \mathbb Q.
\ee
Hence the points A and C both lie on a discretised Riemann Sphere with pole at B. The non-commutativity of spin observables in QM can now be framed as follows: what prevents $SG_B$ and $SG_C$ being counterfactually swapped, keeping the quantum system fixed? That is to say, although in reality the particle was sent from $SG_A$ to $SG_B$ and then to $SG_C$, what's to prevent a counterfactual world where everything is the same, except that the quantum system is sent from $SG_A$ to $SG_C$ and then to $SG_B$? For this counterfactual world to be simultaneously well-defined, we require, in addition to (\ref{ratSG2}) and (\ref{ratSG3}),
\be
\boldsymbol A \cdot \boldsymbol C \equiv \cos \theta_{AC} \in \mathbb Q
\ee
Hence our question can be framed as follows: what's to stop the cosines of the angular lengths of all three sides of the spherical triangle $\triangle ABC$ in Fig \ref{triangle} from being rational numbers, if, in addition, the internal angles are also rational (in degrees)? ITC prevents this, as discussed already. 

Hence, with absolutely overwhelming probability, we have our contradiction: the angular lengths of the three sides of the triangle can't all have rational cosines if the internal angles are rational in degrees.  Hence the counterfactual world where $SG_B$ and $SG_C$ are swapped, keeping everything else fixed, cannot be well defined, simultaneous with the real world, even though the experimenter was free to choose measurement orientations as they liked. That is to say, in our deterministic model with rational number constraints, the non-commutativity of observables is implied by Niven's Theorem.

\subsection{Bell's Theorem}   
\label{bell}

We analyse Bell's inequality \cite{Bell:1964} 
\be
\label{bellineq}
1 \ge
|Co(\boldsymbol A_{\mathrm{nom}}, \boldsymbol B_{\mathrm{nom}}) -Co(\boldsymbol A_{\mathrm{nom}}, \boldsymbol C_{\mathrm{nom}})| -Co(\boldsymbol B_{\mathrm{nom}},\boldsymbol C_{\mathrm{nom}})
\ee
using ITC. Analysing the CHSH inequality is fundamentally no different \cite{Palmer:2024} though a little more complicated. In (\ref{bellineq}), $Co$ denotes a correlation over many experimental runs of individual quantum systems, each prepared in the same singlet state (\ref{singlet}), each run associated with a specific $\lambda$.  

Experimentally, the correlations in (\ref{bellineq}) are determined from measurements on three separate sub-ensembles of particles. Since, as above, experimenters can only set measuring orientations to nominal accuracy, the exact settings $\boldsymbol A$, $\boldsymbol B$ and $\boldsymbol C$ will not be the same for each run of a given sub-ensemble. For the $i$th run in the sub-ensemble where the first correlation in (\ref{bellineq}) is being experimentally found, RaQM requires that the exact settings satisfy:
\be
\label{e1}
\boldsymbol A_i \cdot \boldsymbol B_i \equiv \cos \theta_{A_iB_i} \in \mathbb Q
\ee
Similarly, for the $i$th run in the sub-ensemble where the second correlation of (\ref{bellineq}) is being experimentally found:
\be
\label{e2}
\boldsymbol A_i \cdot \boldsymbol C_i \equiv \cos \theta_{A_iC_i} \in \mathbb Q
\ee
Finally, for the $i$th run in the sub-ensemble where the third correlation of (\ref{bellineq}) is being experimentally found:
\be
\label{e3}
\boldsymbol B_i \cdot \boldsymbol C_i \equiv \cos \theta_{B_iC_i} \in \mathbb Q.
\ee
Since (\ref{e1}), (\ref{e2}) and (\ref{e3}) correspond to separate sub-ensembles of particles, these rationality conditions can be trivially satisfied. With this, RaQM predicts
\begin{align}
Co(\boldsymbol A_{\mathrm{nom}}, \boldsymbol B_{\mathrm{nom}})&= -\boldsymbol A_{\mathrm{nom}} \cdot \boldsymbol B_{\mathrm{nom}},  \nonumber \\
Co(\boldsymbol A_{\mathrm{nom}}, \boldsymbol C_{\mathrm{nom}})&= -\boldsymbol A_{\mathrm{nom}} \cdot \boldsymbol C_{\mathrm{nom}}, \nonumber \\
Co(\boldsymbol B_{\mathrm{nom}}, \boldsymbol C_{\mathrm{nom}})&= -\boldsymbol B_{\mathrm{nom}} \cdot \boldsymbol C_{\mathrm{nom}}
\end{align}
exactly as in QM. Consistent with this, and as found experimentally, Bell's inequality can be violated for suitable choices of nominal measurement setting. 
 
What property of RaQM negates the proof that RaQM must necessarily satisfy (\ref{bellineq})? For a putative local hidden-variable model to satisfy Bell inequalities, we integrate (or sum)
\be
\label{bellsum}
|S(\lambda, \boldsymbol A)S(\lambda, \boldsymbol B)- S(\lambda, \boldsymbol A)S(\lambda, \boldsymbol C)|-S(\lambda, \boldsymbol B)S(\lambda, \boldsymbol C)
\ee
over $\lambda$, for the associated local deterministic spin functions $S(\lambda, \boldsymbol A)$ etc. But it is only possible to perform this sum if all terms in (\ref{bellsum}) are mathematically well defined, for each $\lambda$. Here, as discussed in Section \ref{raqm}, the permutation operator $\xi$ acts as a hidden variable $\lambda$. 

We now refer to Fig \ref{triangle}. For a given run (i.e. fixed $\lambda$) of a Bell experiment suppose Alice chose $\boldsymbol A_{\mathrm{nom}}$ and Bob $\boldsymbol B_{\mathrm{nom}}$, with corresponding exact settings $\boldsymbol A$ and $\boldsymbol B$. Then, in order for (\ref{bellsum}) to be well defined, it must be the case that the two following counterfactual worlds \emph{simultaneously} have well-defined measurement outcomes with the real world: one where Alice's exact setting continues to be $\boldsymbol A$ and Bob chooses $\boldsymbol C_{\mathrm{nom}}$ with some exact but unknown setting $\boldsymbol C$, and one where Bob's exact setting continues to be $\boldsymbol C$ and Alice's exact setting is $\boldsymbol B$. Hence, in order for (\ref{bellsum}) to be well defined, we require that, simultaneously:
\be
\label{ratBell}
\cos \theta_{AB} \in \mathbb Q;\ \ 
\cos \theta_{AC} \in \mathbb Q;\ \ 
\cos \theta_{BC} \in \mathbb Q
\ee
Since $C$ is arbitrary within the neighbourhood defined by $\boldsymbol C_{\mathrm{nom}}$,  the first counterfactual (the second of (\ref{ratBell})) can be satisfied trivially. 

Crucially, the rationality conditions cannot now be satisfied for the second counterfactual. To see this, note that although $\boldsymbol  A_{\mathrm{nom}}$, $\boldsymbol  B_{\mathrm{nom}}$ and $\boldsymbol  C_{\mathrm{nom}}$ are coplanar, the corresponding exact angles $\phi_A$, $\phi_B$ and $\phi_C$ will not be \emph{precisely} equal to $0^\circ$ or $180^\circ$. In addition to (\ref{ratBell}), we must additionally demand that $\phi_A$ (only approximately $0^\circ$ or $180^\circ$) is rational in degrees. But now ITC can again be invoked to imply that it is impossible for the quantum state to be simultaneously well defined in the real-world basis and  \emph{both} of counterfactual-world bases. Of course this conclusion applies no matter what the real-world nominal measurement directions are (i.e., they could be $\boldsymbol  A_{\mathrm{nom}}$ and $\boldsymbol  C_{\mathrm{nom}}$, or $\boldsymbol  B_{\mathrm{nom}}$ and $\boldsymbol  C_{\mathrm{nom}}$). Hence, (\ref{bellsum}) is generally undefined. Neither realism nor locality are violated here. Rather, the Measurement Independence assumption (MI) is violated \cite{HossenfelderPalmer}, not for the freely-chosen nominal settings, but for the exact settings that were never under the experimenters' control anyway \cite {Palmer:2026a}. More specifically, 
\be
\label{MI}
\rho(\lambda | \boldsymbol A_{\mathrm{nom}},  \boldsymbol B_{\mathrm{nom}}) = \rho(\lambda)
;\ \ \ 
\rho(\lambda | \boldsymbol A,  \boldsymbol B) \ne \rho(\lambda)
\ee

\section{Why Nature Makes Use of Complex Numbers}
\label{complex}

Complex numbers are an essential feature of QM, and indeed it is claimed that quantum theory based on real numbers can be experimentally falsified \cite{Renou:2021}. But why?

QM is itself based on the field $\mathbb C$, where the incomprehensible $i=\sqrt 2$ must be introduced axiomatically. It is certainly mysterious - if not miraculous - that such a quantity should play a vital role in physics as remarked in many introductory texts on QM. One way of justifying such quantities (e.g. \cite{Penrose:2004}) is to note that extending the rationals to include the irrationals works well in classical physics (where, as concerned the ancient Greeks, $\sqrt 2$ is also rather incomprehensible). Hence, there is no reason to suppose that extending the reals further by adjoining the further incomprehensible $i$ shouldn't work in transitioning from classical to quantum physics. That is to say, the same trick that worked once will surely work again. 

Of course, in classical physics there is no \emph{need} to extend the rationals to the reals. The differential equations of classical physics are (typically) sufficiently well behaved that it can be proved that numerical solutions converge to the exact solutions as discretisations go to zero \cite{Strogatz}. And in RaQM we do not extend the rationals to the irrational reals in any case. We have argued that, in quantum physics at least, this extension is profoundly obfuscating - it hides number theoretic properties which are actually quite important. Which is to say, the trick does not work for the irrational reals, and hence shouldn't work again when we extend the reals to the complexes. And this is indeed what is found. In RaQM $i$ is not some unphysical quantity only defined by axiom, it is instead defined constructively as a permutation/negation operator where $i\{a_1, a_2\}=\{-a_2, a_1\}$, and $a_i \in \{1,-1\}$.

That is to say, in answering the question `why complex numbers?', we should distinguish \emph{complex structure}, leading to the arithmetically incomplete discretised Riemann Sphere, from the arithmetically complete field $\mathbb C$. As discussed, the incomplete gappy nature of the Riemann Sphere allows us to interpreted the violation of Bell's Inequality as confirming the holistic nature of the laws of physics (already educed by an analysis of single-particle interference). Without these gaps, i.e. in the singular limit where the discretised Riemann Sphere goes to $\mathbb C \cup \{\infty\}$, RaQM would have the property of counterfactual completeness and then one would be forced into interpreting the violation of Bell inequalities as a violation of local realism, which, as Einstein consistently argued, is physically unacceptable. 

The discretised Riemann Sphere is essential to constructing our gappy Hilbert Space. We noted, for example, that the construction encompasses higher-order quaternionic multiplication, consistent with discretised rotations in 3D physical space. Indeed the construction is consistent with Lorentz invariance (and hence the Dirac Equation), an issue which will be explored elsewhere. We therefore conclude that the reason nature makes use of complex structure (but not the continuum field $\mathbb C$) is that the laws of physics are fundamentally holistic, some examples being discussed below. 

\section{Conclusions} 
\label{conclusion}

The author has developed a theory of quantum physics, named Rational Quantum Mechanics (RaQM), where complex Hilbert Space is discretised. It is proposed that discretisation is due to gravity. The weakness of gravity implies that for many purposes RaQM and QM agree on their predictions. Indeed, one could go far as to say the weakness of gravity is the reason why QM has been such a successful theory. This has rather profound consequences for approaches to synthesise quantum and gravitational physics. Instead of thinking of gravity as some `force' to be quantised like the other forces of nature, gravity plays an inherent role in determining the structure of basic quantum physics. On this basis, it perhaps it shouldn't be so surprising if it is found that vacuum energy doesn't gravitate, a topic that will be discussed in more detail elsewhere. 

Why should anyone believe in RaQM? In \cite{Palmer:2026b}, the author has proposed an experiment where QM and RaQM will have testable differences. It is that the exponential advantage of quantum algorithms that utilise the quantum fourier transform, will cease in quantum computers which utilise more than 1,000 perfect (i.e. logical) qubits.

However, it may be a few years before such an experiment can be performed. In this paper, therefore, we discuss a rather different argument for why one should believe in RaQM: essentially by eschewing the continuum, RaQM solves all of the mysteries of QM - from the basic mystery of interference to the puzzling implications of the violation of Bell's inequality. Solving these mysteries indicates that whilst the laws of physics are profoundly holistic, they are not nonlocal. 

It is worth describing two profoundly holistic, but not nonlocal, concepts. The first is Mach's Principle, that inertia here is due to mass there. No-one suggests that the mass `there' is somehow acting nonlocally. And indeed Mach's Principle was a major motivation for Einstein in formulating his theory of general relativity. The second is the geometry of the fractal attractor of a deterministic dynamical system. Although the dynamical equations of such a system are classical, the attractor is itself an emergent asymptotic property and arguably not itself classical (for example, it is formally non-computable). As the author has speculated with his Invariant Set Postulate \cite{Palmer:2009a}, it is possible that some fractal state space geometry belonging to the dynamics of the universe as a whole describes the holistic underpinning to RaQM. The equations describing such a geometry would have to be $p$-adic in nature, $p$-adic numbers being to fractal geometry as real numbers are to Euclidean geometry. As we have already discussed, the appropriate metric for state space in RaQM is the $p$-adic metric. This prevents RaQM from being considered a fine-tuned theory.  

Our analysis has finally revealed why nature makes use of complex numbers. The discretisation of the Bloch Sphere, where the discretisation along a line of longitude is incommensurate with the discretisation along a line of longitude (the incommensurateness being described by Niven's Theorem), inevitably involves complex structure. However, importantly, it is claimed that nature does not make use of the field $\mathbb C$ of complex numbers, and hence of an object $i$ only definable by axiom. In RaQM, $i$ is simply a permutation/negation operator acting on 2-bit strings. 

Finally, one may ask why this paper concurs with Feynman's claim that interference is the only mystery, when Spekkens \cite{Spekkens:2007} claims otherwise. It is because, the author suggests, the proposal that `nature abhors a continuum' digs deeper into the core principles of fundamental physics than does Spekkens Knowledge-Balance Principle (KBP) - `that if one has maximal knowledge, then for every system at every time, the amount of knowledge one has about the ontic state of the system at that time must equal the amount of knowledge one lacks'. Spekkens finds KBP is enough to explain interference but not Bell's Theorem. By our analysis of complementarity for example (see Section \ref{complementarity}), it can be seen that RaQM embraces KBP. But RaQM's axioms transcend KBP. From RaQM's perspective, Feynman was right, as indeed he usually was. 

\bibliography{mybibliography}

\begin{thebibliography}{10}

\bibitem{Bell:1964}
J.S. Bell.
\newblock On the {E}instein-{P}odolsky-{R}osen paradox.
\newblock {\em Physics}, 1:195--200, 1964.

\bibitem{Berry}
M~Berry.
\newblock Singular limits.
\newblock {\em Physics Today}, 55:10--11, 2002.

\bibitem{EPR}
A.~Einstein, B.~Podolsky, and N.~Rosen.
\newblock Can quantum-mechanical description of physical reality be considered
  complete.
\newblock {\em Physical Review}, 47:777--780, 1935.

\bibitem{Ellis:2018}
G.F.R. Ellis, K.A.Meissner, and H.Nicolai.
\newblock The physics of infinity.
\newblock {\em Nature}, 14:770--772, 2018.

\bibitem{Feynman}
R.P. Feynman, R.B.Leighton, and M.Sands.
\newblock {\em The Feynman Lectures on Physics, Vol III}.
\newblock Basic Books, 2011.

\bibitem{HossenfelderPalmer}
Sabine Hossenfelder and Tim Palmer.
\newblock Rethinking superdeterminism.
\newblock {\em Frontiers in Physics}, 8:139, 2020.

\bibitem{Jahnel:2005}
J.~Jahnel.
\newblock When does the (co)-sine of a rational angle give a rational number?
\newblock arXiv:1006.2938, 2010.

\bibitem{Niven}
I.~Niven.
\newblock {\em Irrational Numbers}.
\newblock The Mathematical Association of America, 1956.

\bibitem{Palmer:2024}
Tim Palmer.
\newblock Superdeterminism without conspiracy.
\newblock {\em Universe}, 10:47, 2024.

\bibitem{Palmer:2026a}
Tim Palmer.
\newblock Impossible counterfactuals, discrete {H}ilbert space and {B}ell's
  theorem.
\newblock {\em Journal of Physics - Conference Series. Issue in Memory of Basil
  Hiley}, arXiv:2601.14941, 2026.

\bibitem{Palmer:2026b}
Tim Palmer.
\newblock Rational quantum mechanics: testing quantum theory with quantum
  computers.
\newblock {\em Proceedings of the National Academy of Sciences},
  arXiv:2601.14941, 2026.

\bibitem{Palmer:2009a}
T.N. Palmer.
\newblock The invariant set postulate: a new geometric framework for the
  foundations of quantum theory and the role played by gravity.
\newblock {\em Proc. Roy. Soc.}, A465:3165--3185, 2009.

\bibitem{Penrose:2004}
R.~Penrose.
\newblock {\em The Road to Reality: A Complete Guide to the Laws of the
  Universe}.
\newblock Jonathan Cape, London, 2004.

\bibitem{Renou:2021}
M.O. Renou, D.~Trillo, and M.~Weilenmann et~al.
\newblock Quantum theory based on real numbers can be experimentally falsified.
\newblock {\em Nature}, 600:625--629, 2021.

\bibitem{Schwinger}
J.~Schwinger.
\newblock {\em Quantum Mechanics: Symbolism of Atomic Measurements}.
\newblock Springer, 2001.

\bibitem{Spekkens:2007}
R.W. Spekkens.
\newblock Evidence for the epistemic view of quantum states: {A} toy theory.
\newblock {\em Phys. Rev. A}, A75:032110, 2007.

\bibitem{Strogatz}
S.H. Strogatz.
\newblock {\em Nonlinear dynamics and chaos}.
\newblock Westview Press, 2000.

\bibitem{Wheeler}
J.~Wheeler.
\newblock Information, physics, quantum: the search for links.
\newblock {\em 3rd Int Symp. Foundations of Quantum Mechanics, Tokyo}, pages
  354--368, 1989.

\end{thebibliography}

\section*{Appendix}

\section*{Discretising the Riemann Sphere}
\label{Riemann}

The continuum field $\mathbb C$, with its axiomatically defined element $i=\sqrt{-1}$, plays an essential role in QM (either explicitly or implicitly as in the de Broglie-Bohm interpretation). Central to RaQM is a discretisation of the Riemann Sphere $\mathbb C \cup \{\infty\}$ of (extended) complex numbers $z(\theta, \phi)=\cot (\theta/2)e^{i \phi}$ where $\theta$ denotes co-latitude (or zenith angle) and $\phi$ denotes longitude (or azimuth). The discretisation is defined by
\begin{equation}
\label{ratL}
\cos^2 \frac{\theta}{2}= \frac{m}{L} \in \mathbb Q ; \ \ \ \ \frac{\phi}{2\pi}= \frac{n}{L} \in \mathbb Q.
\end{equation}
where $L \in \mathbb N$ and $0 \le m, n \le L$ are whole numbers. Note that if $\theta$ satisfies (\ref{ratL}) then since $\cos \theta = 2 \cos^2 (\theta/2)-1$, necessarily $\cos \theta \in \mathbb Q$. $L$ is an especially important variable, defining the degree of granularity of discretised Hilbert Space: QM is the (singular) limit of RaQM at $L=\infty$, whilst the classical limit of RaQM occurs at $L=1$. The key property of this discretisation is that the $i$ (with property $i^2=-1$) does not have to be introduced by axiom. Rather, as discussed below, $i$ and the corresponding quaternions are defined constructively from representations of complex numbers as permutation/negation operators acting on bit strings. With this definition of $i$, the entire set $\{z(\theta, \phi)\}$ of complex numbers on the discretised Riemann Sphere is obtained constructively by an inductive argument, for all $L \in \mathbb N$. 

\subsection*{Discretised Complex Numbers and Quaternions}

When $L=2^0=1$, (\ref{ratL}) implies $\theta \in  \{0, \pi\}$, $\phi =0$, i.e., the discretisation only admits 2 points at the north and south poles. We represent the corresponding set of `complex numbers' in terms of the identity and negation operators
\be
\label{L1}
z(0, 0) \{1\}=\{1\};\ \ z(\pi, 0)\{1\}=\{-1\}=-\{1\}
\ee
With $L=2^0$ (the classical limit of RaQM), the discretisation is too coarse to admit complex structure at all. Complex numbers play no essential role in classical physics. 

\begin{figure}
\centering
\includegraphics[scale=0.65]{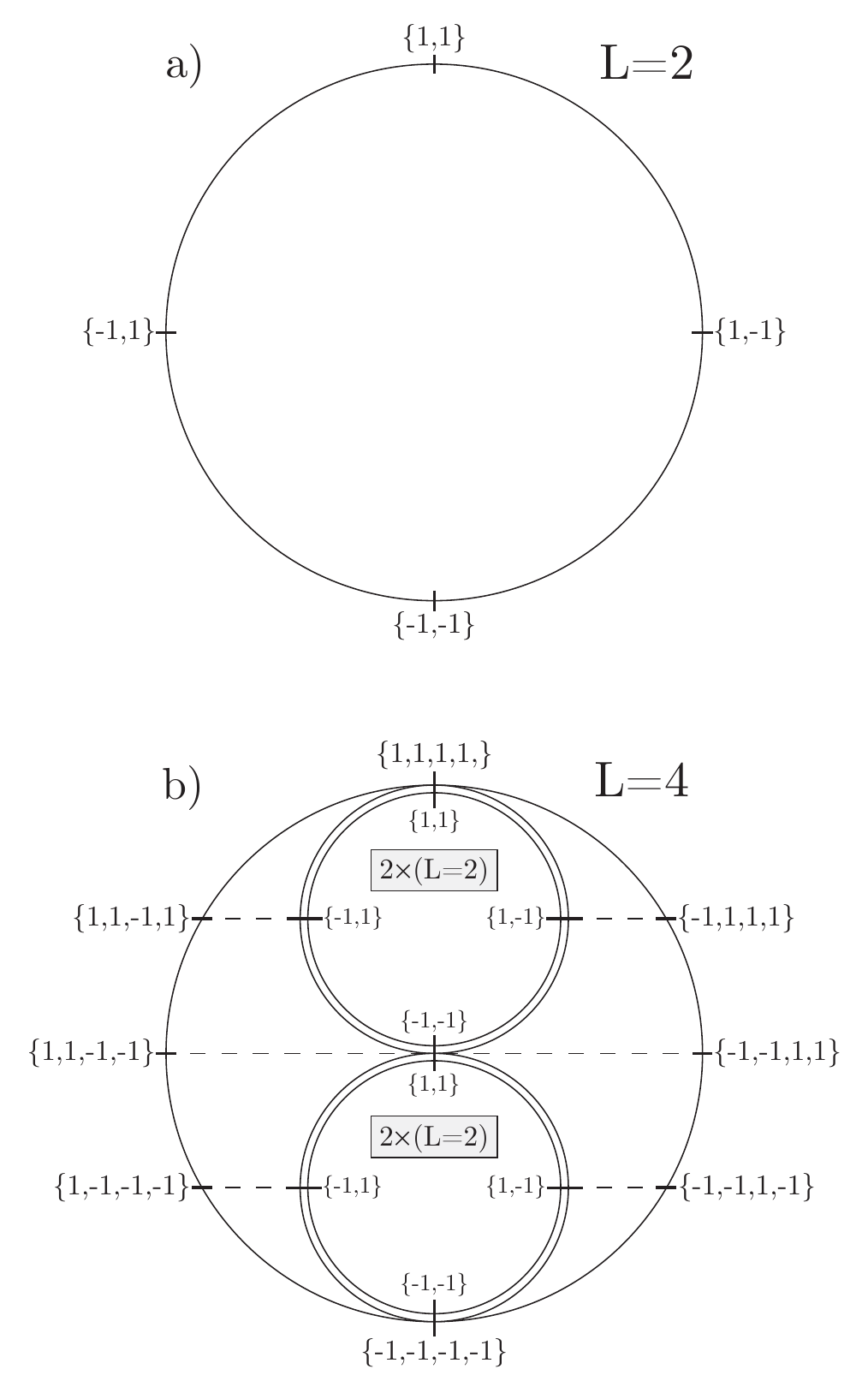}
\caption{\emph{a) The discretised Riemann Sphere at $L=2$, represented as four 2-bit strings on a complete great circle at $\phi=0,\pi$ through the poles. b) A schematic illustration of how to construct the discretised Riemann Sphere for $L=4$, again on the great circle through $\phi=0,\pi$, from two $L=2$ great circles as in a). See the text for details. This also illustrates the construction of an $L=2^M$ discretised Riemann Sphere on the great circle at $\phi=0, \pi$, from two $L/2$-bit discretised great circles. }}
\label{Riemann1}
\end{figure}

When $L=2^1$, then $\theta \in \{0, \pi/2, \pi\}$, $\phi \in \{0, \pi\}$ according to (\ref{ratL}). $L=2$ is large enough to admit complex structure, represented by the permutation/negation operator (PNO) $i$ defined as
\be
\label{i}
 i\{a_1,a_2\}=\{-a_2, a_1\}
 \ee
where $a_i \in \{1, -1\}$, so that $i^2\{a_1,a_2\}=\{-a_1, -a_2\} \equiv -\{a_1, a_2\}$. Generalising (\ref{L1}), the set of complex numbers on the $L=2$ discretised Riemann Sphere are associated with the 4 PNOs $\{i^0, i, i^2, i^3\}$. Specifically, 
\begin{align}
&z(0, 0)\{1,1\}=i^0\{1,1\}=\{1,1\}; \nonumber \\
&z(\frac{\pi}{2}, 0)\{1,1\}=i\{1,1\}=\{-1,1\}; \nonumber \\
&z(\pi, 0)\{1,1\}=z(\pi,\pi)\{1,1\}=i^2 \{1,1\}=\{-1,-1\}; \nonumber \\
&z(\frac{\pi}{2}, \pi)\{1,1\}=i^3\{1,1\}=\{1,-1\}
\end{align}
as shown in Fig \ref{Riemann1}a.  

When $L=2^2=4$ then $\theta \in \{0, \pi/3, \pi/2, 2\pi/3, \pi\}$, $\phi \in \{0, \pi/2, \pi, 3\pi/2\}$ according to (\ref{ratL}). The discretised sphere now comprises 8 points on each of two complete great circles (at $\phi=0, \pi$ and $\phi=\pi/2, 3\pi/2$) passing through the north and south poles - 14 points in total. $L=4$ is now large enough to admit quaternionic structure, as shown below. The complex numbers at these points are represented as PNOs acting on 4-bit strings. We start by considering the points on the great circle $\phi=0,\pi$, writing $\{1,1,1,1\}$ as two concatenated 2-bit strings $\{1,1\} |\{1,1\}$. We define PNOs based on $\{i^0, i, i^2, i^3\}$ acting individually on each pair of 2-bit strings, with a construction motivated by the theory of spinors. Specifically, we first apply $i$ twice to the second 2-bit string keeping the first fixed. We then apply $-i$ twice to the first 2-bit string keeping the second fixed; then again apply $i$ twice to the second bit string keeping the first fixed; then again apply $-i$ twice to the first bit string keeping the second fixed. In total, from Fig \ref{Riemann1}a, the first 2-bit string has been rotated by $4\pi$ relative to the second (like Dirac scissors relative to the back of the chair). Specifically, 
\begin{align}
z(0, 0)\{1,1,1,1\}=&\; i^0\{1,1\}|i^0 \{1,1\}=\{1,1,1,1\}; \nonumber \\
z(\frac{\pi}{3}, 0)\{1,1,1,1\}=&\;  i^0\{1,1\} | i^1 \{1,1\}=\{1,1,-1,1\};\nonumber \\
z(\frac{\pi}{2}, 0)\{1,1,1,1\}=&\; i^0 \{1,1\} | i^2 \{1,1\}=\{1,1,-1,-1\};\nonumber\\
z(\frac{2\pi}{3}, 0)\{1,1,1,1\}=&\; i^3 \{1,1\} | i^2 \{1,1\}=\{1,-1,-1,-1\}; \nonumber \\
z(\pi,0)\{1,1,1,1\}=z(\pi,\pi)\{1,1,1,1\}=&\;  i^2 \{1,1\} | i^2 \{1,1\}=\{-1,-1,-1,-1\}; \nonumber \\
z(\frac{2\pi}{3}, \pi)\{1,1,1,1\}=&\;  i^2 \{1,1\} | i^3 \{1,1\}=\{-1,-1,1,-1\};\nonumber \\
z(\frac{\pi}{2},\pi)\{1,1,1,1\}=&\;  i^2 \{1,1\} |\ i^0 \{1,1\}=\{-1,-1,1,1\};\nonumber \\
z(\frac{\pi}{3},\pi)\{1,1,1,1\}=&\; i^1 \{1,1\} | i^0 \{1,1\}=\{-1,1,1,1\}
\end{align}
This construction is illustrated schematically in Fig \ref{Riemann1}b. Here the 4-bit strings on the $L=4$ great circle comprise two 2-bit strings from the pair of smaller $L=2$ circles, where one rotates in steps of $\pi/2$ by $4 \pi$ relative to the other. 

\begin{figure}
\centering
\includegraphics[scale=0.8]{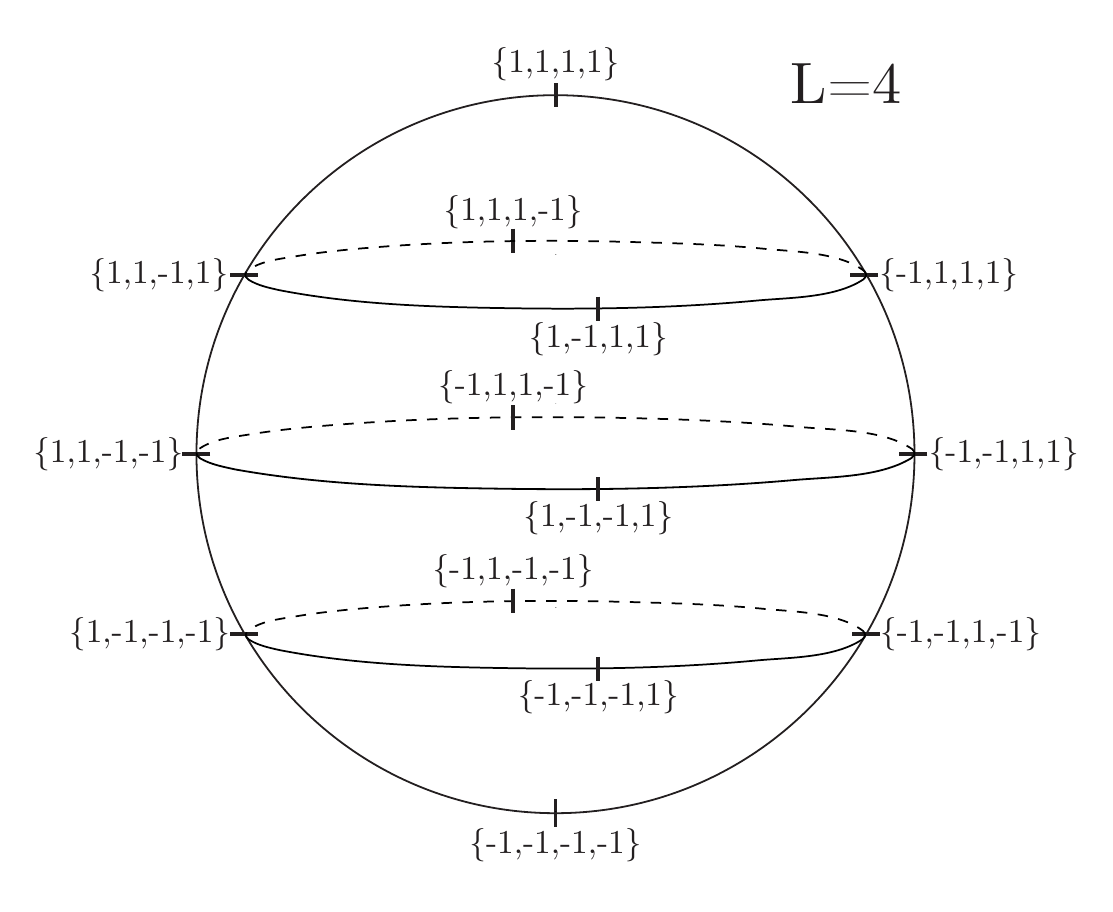}
\caption{\emph{The full 3D discretised Riemann Sphere for $L=4$ - where, consistent with quaternionic structure (see text), rotation about the polar axes corresponding to a cyclic permutation of the 4-bit strings.}}
\label{Riemann2}
\end{figure}

The complex numbers on the $\phi=\pi/2, 3\pi/2$ great circle are PNOs obtained by a cyclic permutation of bit strings on the $\phi =0, \pi$ great circle, so that
\begin{align}
z(0, \frac{\pi}{2})\{1,1,1,1\}=&\; \{1,1,1,1\}; \nonumber \\
z(\frac{\pi}{3}, \frac{\pi}{2})\{1,1,1,1\}=&\; \{1,-1,1,1\}; \nonumber \\
z(\frac{\pi}{2}, \frac{\pi}{2})\{1,1,1,1\}=&\; \{1,-1,-1,1\};\nonumber\\
z(\frac{2\pi}{3}, \frac{\pi}{2})\{1,1,1,1\}=&\; \{-1,-1,-1,1\}; \nonumber \\
z(\pi,\frac{\pi}{2})\{1,1,1,1\}=z(\pi,\frac{3\pi}{2})=&\; \{-1,-1,-1,-1\}; \nonumber \\
z(\frac{2\pi}{3}, \frac{3\pi}{2})\{1,1,1,1\}=&\; \{-1,1,-1,-1\}; \nonumber \\
z(\frac{\pi}{2},\frac{3\pi}{2})\{1,1,1,1\}=&\; \{-1,1,1,-1\}; \nonumber \\
z(\frac{\pi}{3},\frac{3\pi}{2})\{1,1,1,1\}=&\; \{1,1,1,-1\}.
\end{align}
That this cyclic permutation is consistent with quaternionic structure can be seen by defining the operators
\begin{align}
I \{a_1, a_2, a_3, a_4\} =  \{a_3, a_4, -a_1, -a_2\} \nonumber \\
J \{a_1, a_2, a_3, a_4\} =  \{a_2, -a_1, -a_4, a_3\} \nonumber \\
K \{a_1, a_2, a_3, a_4\} =  \{-a_4, a_3, -a_2, a_1\} 
\end{align}
which satisfy 
\begin{equation}
I^2=J^2=K^2=- 1; \ \ \ I \times J=K.
\end{equation}
In particular, $I\{1,1,1,1\}=\{1,1,-1,-1\}=z(\pi/2,0)\{1,1,1,1\}$ and $J \{1,1,1,1\}=\{1,-1,-1,1\}=z(\pi/2, \pi/2)\{1,1,1,  1\}$, consistent with $I$ and $J$ rotating the 4-bit string $\{1,1,1,1\}$ at the north pole to the 4-bit strings on the equator at $\phi=0$ and $\phi=\pi/2$, respectively. Consistent with this, $K$ maps one these equatorial points to the other: 
\be
KI\{1,1,1,1\}=K\{1,1,-1,-1\}=\{1,-1,-1,1\}=J\{1,1,1,1\}.
\ee
The 14 4-bit strings on the 3D $L=4$ discretised Riemann sphere are shown in Fig \ref{Riemann2}.
The inductive step is to write the discretised PNOs for some $L=2^M$, acting on the $L$-bit string $\{1,1,\ldots, 1\}$ as PNOs acting on 2 concatenated $L/2$-bit strings $\{1,1,\ldots 1\}|\{1,1,\ldots, 1\}$. The PNOs on the $\phi=0, \pi$ great circle are defined using the spinorial construction illustrated in Fig \ref{Riemann2}b, generalised so that the two smaller circles represent the two $L/2$ discretised great circles at $\phi=0, \pi$. We then rotate the $\phi=0, \pi$ great circle around the polar axis by cyclically permuting the $L$-bit string as in 
\be
\label{zeta}
\zeta \{a_1, a_2, \ldots, a_L\} \mapsto \{a_2, a_3, \ldots, a_L, a_1\}
\ee
Each application of $\zeta$ corresponds to a rotation by $2\pi/L$ radians. 

This completes the $L$-discretisation of the Riemann Sphere when $L$ is a power of 2. For discretisations where $L$ is not a power of 2, one can simply interpolate between constructions where $L$ is a power of 2. For example, with $L=3$, one can simply remove the last bit from the $L=4$ construction on $\phi=0$, yielding
\begin{align}
&z(0, 0)\{1,1,1\}=\{1,1,1\}; \nonumber \\
&z(\theta_*, 0)\{1,1,1\}=\{1,1,-1,\};\nonumber \\
&z(\pi-\theta_*, 0)\{1,1,1\}=\{1,-1,-1\}; \nonumber \\
&z(\pi,0)\{1,1,1\}=\{-1,-1,-1\}
\end{align}
where $\cos \theta_*=1/3$. For the longitudes $\phi=2\pi/3, 4\pi/3$, one simply performs two cyclic permutations to the 3-bit strings. 

Note that, for all $L$, the proportion of 1s in the bit-string representing $z(\theta, \phi)$ is equal to $\cos^2 \theta/2$. Hence the proportion of $-1$s equals $\sin^2 \theta/2$, and the relative frequency of $1$s to $-1$s is equal to $\cot^2 \theta/2$. This is the basis for the claim that there is no need to postulate Born's Rule in RaQM - it is redundant in the Dirac-von Neumann axioms.  

 \end{document}